\documentclass[pre,twocolumn,floatfix,superscriptaddress,showpacs,showkeys,preprintnumbers,amsmath,amssymb]{revtex4}

\usepackage{graphicx}
\usepackage{dcolumn}
\usepackage{bm}

\begin{document}

\preprint{APS/123-QED}

\title{Bifurcation and Chaos in Coupled Ratchets \\exhibiting
            Synchronized Dynamics } 
\author{U.~E.~Vincent}
\affiliation{Department of Physics, College of Natural Sciences,
  University of Agriculture, Abeokuta, Nigeria.}
\affiliation{Department of Physics and Solar Energy, Bowen University, Iwo, Nigeria.}
\author{A. Kenfack}
\email[Corresponding author:]{kenfack@mpipks-dresden.mpg.de}
\affiliation{Max Planck Institute for the Physics of Complex Systems, N$\ddot
  o$thnitzer Strasse 38, 01187 Dresden, Germany.}
\author{A.~N.~Njah}
\affiliation{Department of Physics, College of Natural Sciences, University of Agriculture, Abeokuta, Nigeria.}
\author{O.~Akinlade}
\affiliation{Department of Physics, College of Natural Sciences, University of Agriculture, Abeokuta, Nigeria.}

\date{\today}

\begin{abstract}
The bifurcation and chaotic behaviour of unidirectionally coupled deterministic
ratchets is studied as a function of the driving force amplitude ($a$) and
frequency ($\omega$). A classification of the various types of bifurcations
likely to be encountered in this system was done by examining the stability of
the steady state in linear response as well as constructing a two-parameter phase
diagram in the ($a -\omega$) plane. Numerical explorations revealed varieties of
bifurcation sequences including quasiperiodic route to chaos. Besides, the familiar period-doubling and crises route to chaos exhibited
by the one-dimensional ratchet were also found. In addition, the coupled ratchets
display symmetry-breaking, saddle-nodes and bubbles of bifurcations. Chaotic
behaviour is characterized by using the sensitivity to initial condition as well
as the Lyapunov exponent spectrum; while a perusal of the phase space
projected in the Poincar$\acute{e}$ cross-section confirms some of the
striking features.
\end{abstract}

\maketitle
\section{Introduction}
Coupled nonlinear systems possess a rich catalog of potentially useful
dynamical behaviours including bifurcations, chaos and
synchronization. They are central to the understanding of a wide
variety of extended systems, e.g. a line of lattice oscillations or
coupled plasma wave modes. A system of two oscillators could make a
Hopf bifurcation to a second incommensurate frequency and follow a
quasiperiodic route to chaos, in addition to period doubling and
intermittency.  Such routes to chaos was first observed by Ruelle and
Takens\cite{ruelle71}.
Synchronization phenomena in coupled or driven nonlinear oscillators
are of fundamental importance and have been extensively investigated
both theoretically and experimentally in the context of many specific
problems arising in laser dynamics, electronic circuits, secure communications
and time series analysis \cite{zhang02} to mention a few. In a recent study,
we observed phase synchronization in both unidirectionally \cite{vincent04} and
bidirectionally \cite{vincent05} coupled deterministic ratchets that exhibits
intermittent chaos.  In \cite{vincent04,vincent05}, it was shown that the transition to
the synchronous regime is characterized by an interior crises transition
of the attractor in the phase space.  Specifically, two analytic tests
based on the Fujisaka and Yamada approach \cite{fujisaka83,yamada83} and that of
Gauthier and Bienfang \cite{gauthier96} were employed to verify the stability
of the synchronous state in reference \cite{vincent04}.

A clear understanding of coupled oscillator systems requires a study
of the full spectrum of the operating regimes, including the nonsynchronous
state \cite{ram00}. It has been reported that a group of interacting
chaotic units that exhibits synchronization show some bifurcation
cascades from disorder to partial and global order when varying the
coupling strength \cite{zhang00}. Ding and Yang \cite{ding96} showed the
existence of intermingled basins in coupled Duffing oscillators that
exhibit synchronized chaos and conjectured that intermingled basins
can be easily realized in the context of coupled oscillators and
synchronized chaos.  Extensive bifurcation analysis of two coupled
periodically driven Duffing oscillators was done by Kozlowski et
al. \cite{kozlowski95}. They showed that the global pattern of bifurcation
curves in parameter space consists of repeated subpatterns similar to
the superstructure observed for single, periodically driven, strictly
dissipative oscillators.  This study was recently extended to two
coupled periodically driven double-well Duffing oscillators by Kenfack
\cite{kenfack03}. The results revealed a striking departure from the single-well Duffing oscillators studied by Kozlowski et al \cite{kozlowski95}.

The literature is abundant with studies on coupled oscillator
systems. At this point, we wish to refer the reader to some other
recent studies that are related to the subject of this work
~\cite{wirkus02}-\cite{raj99}. Motivated by this series of studies, we aim in this paper to
explore some dynamical features hitherto not fully explored in the
coupled ratchets with emphasies on the bifurcation structures
preceeding the stable synchronous regime. The rest of the paper is
organized as follows: Section 2 describes the equations of motion of
the chaotic ratchet model and comment on the synchronization behaviour
of the coupled system that is of interest in this study. In Section 3,
we carry out a linear stability analysis of the system projected on
the Poincar$\acute{e}$  map, while in Section 4, numerical results of
bifurcations are presented. Chaotic behaviour is characterized in
Section 5. The paper is concluded in Section 6.

\section{The Chaotic Ratchet Model}
\begin{figure}[h]
\vskip 0.25cm
\includegraphics[width=7cm,height=4cm]{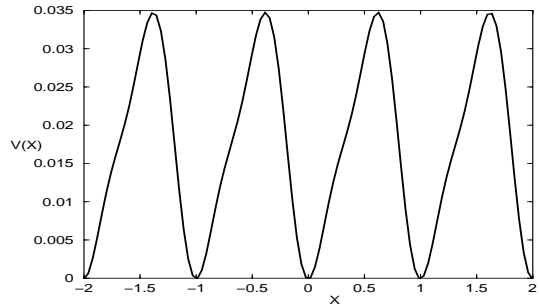}
\caption{\label{figure1} The ratchet potential ($C=0.0173, \delta = 1.6$)}
\end{figure}

Let us consider the one-dimensional problem of a particle driven by a
periodic time-dependent external force under the influence of an
asymmetric potential of the ratchet type~\cite{jung96}-\cite{mateos03b}. The time average of the external force is zero. In the absence of stochastic noise, the
dynamics is exclusively deterministic. The dimensionless equation of
motion for a particle of unit mass moving in the ratchet potential
$V(x)$ is given by
\begin{equation}
{\ddot x + b \dot x + {\frac{dV(x)}{dx}} = a \cos(\omega_{D}t)}
\end{equation}
Where time $t$ has been normalized in the unit of $\omega_{0}^{-1}$,
the small resonant frequency of the system. Here, $a$ and $b$ are the
forcing strength and the damping parameter, respectively. The dimensionless
potential $V(x)$ is given by
\begin{equation}
V(x) = C - \frac{1}{4\pi^{2}\delta}\lbrack\sin2\pi(x-x_{0})+0.25\sin4\pi(x-x_{0})\rbrack,
\end{equation}
with constants $C \simeq 0.0173$ and $\delta \simeq 1.6$. The potential
$V(x)$ is shifted by a value $x_{0}$  in order to place its minimum at
the origin (see fig.~\ref{figure1}). Notice that this asymmetric
potential is periodic and has infinite potential wells.

\begin{figure}
\vskip 0.25cm
\includegraphics[width=7cm,height=5cm,angle=0]{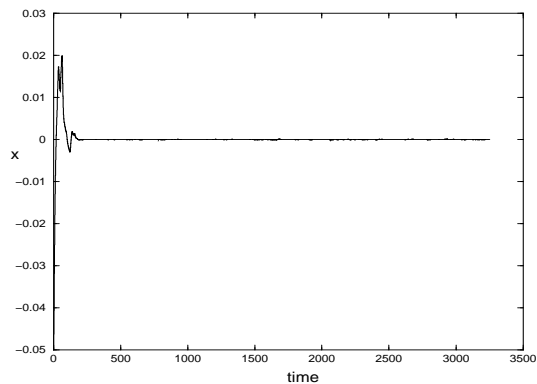}
\caption{\label{figure2} Synchronization dynamics showing the quantity
  $x=x_2-x_1$ as function of time, for $c=0.95$, $b=0.1$,
  $a=0.08092844$ and $\omega=0.67$}
\end{figure}

The extended phase space in which the dynamic is taking place is
three-dimensional, since we are dealing with an inhomogeneous
differential equation with an explicit time dependence.  We can rewrite equation (1) in autonomous form as a three dimensional dynamical system described by the coordinates:
\begin{eqnarray}
        \dot x &=& y, \nonumber \\
        \dot y &=& a \cos z - by - \frac{dV(x)}{dx}, \\
        \dot z &=& \omega_{D}. \nonumber
\end{eqnarray} 
Since equation (3) is nonlinear its solutions allow the possibility
of periodic and chaotic orbits. For the purpose of this study, we
consider a system of two identical unidirectionally coupled chaotic
ratchets governed by:
\begin{eqnarray}
        \ddot x_1 + b \dot x_1 + \frac{dV (x_1)}{d x_1} &=&
         a \cos(\omega t)\\
        \ddot x_2 + b \dot x_2 + \frac{dV (x_2)}{d x_2} &=&
         a \cos(\omega t) + c (\dot x_1- \dot x_2)
\end{eqnarray}
Where $c$ is the coupling parameter. For $a=0.08092844$, $b=0.1$ and
$\omega=0.67$, systems (4) and (5) exhibit stable phase
synchronization when $0.89\le c\le 1.05$. The synchronization
dynamics for $c=0.95$ shown in fig.~\ref{figure2}, is determined by
the behaviour of the following difference $x=x_2-x_1$. The phase
difference is locked (full synchronization, $x=0$) after a short
transient time. Prior to the phase locking event, the strange
attractor undergoes a crises transition during which it is gradually destroyed and as the synchronization regime is
approached, the attractor is again re-built \cite{vincent04}. To obtain
approximate visualization of the attractors and their bifurcations we
follow the procedure described in \cite{kozlowski95,kenfack03}. That is, we investigate
the dynamics in the Poincar$\acute{e}$ sections denoted by $\sum $.

\section{Stability Analysis}
To begin with, let us consider the basic mathematical principle
underlying some expected topological structure of the coupled
ratchets.  Here we carry out a stability analysis of the coupled
system (4) and (5) which can be written equivalently as
\begin{equation}
        \frac{dV}{dt} = F(V,\psi)
\end{equation}
Where  $V(x_1,v_1,x_2,v_2,\omega)$ is an autonomous vector
field and $\psi (a,b,c,\omega)$ is an element of the parameter
space. Thus (4) and (5) become:
\begin{widetext}
\begin{eqnarray}
        \dot x_1 &=& v_{1} \nonumber \\
        \dot v_1 &=& a \cos(\omega t)-bv_1+\frac{1}{4\pi\delta}{\bigg [} 2 \cos2\pi(x_1-x_0)+\cos4\pi(x_1-x_0) {\bigg ]} \nonumber \\
        \dot x_2 &=& v_2 \nonumber \\
        \dot v_2 &=& a \cos(\omega
        t)-bv_2+\frac{1}{4\pi\delta}{\bigg [} 2
        \cos2\pi(x_2-x_0)+\cos4\pi(x_2-x_0) {\bigg ]}+c(v_1-v_2)
\end{eqnarray}
\end{widetext}
The system described by (7) generates a flow $\phi = \{ \phi^{T} \}$
on the phase space $\Re ^{4} \times S^{1}$ such that a global map

\begin{eqnarray}
P: \sum &\to& \sum \nonumber\\
V_{p}(x_1,v_1,x_2,v_2)&\to& P(V_{p}) = \{ \phi^{T} \} |_{\sum (x_1,v_1,x_2,v_2,\theta_0)}\nonumber
\end{eqnarray}
exist, with $\theta_{0}$ being a constant determining the location of
the Poincar$\acute{e}$  cross-section defined by $\sum = \{
(x_1,v_1,x_2,v_2,\theta) \ \epsilon  \  \Re ^{4} \times S^{1}
| \theta = \theta_{0} \}$ on which coordinates of
attractors $(x_1,v_1,x_2,v_2)$ are expressed. By employing the linear perturbation method which consists of
considering the solution $X(x_1,v_1,x_2,v_2)$ as a superposition of a very
small perturbation $Y(\delta x_1,\delta v_1,\delta x_2, \delta v_2)$
to the steady state $V_{0}(x_{10},v_{10},x_{20},v_{20})$, that is
$X=V_0+Y$, we obtain the matrix variational equation
 \begin{equation}
        \dot Y = DG(V_{0})Y
 \end{equation}
where $DG(V_{0})$ is the $4 \times 4$ Jacobian matrix
\begin{equation}
        DG(V_{0})=
\left( \begin{array}{cccc}
        0 & 1 & 0 & 0\\
        \alpha & -b & 0 & 0\\
        0 & 0 & 0 & 1\\
        0 & c & \beta & -(b+c)
\end{array} \right)
\end{equation} 
describing the vector field along the solution $Y$ and $V_0$ being
an equilibrium point or steady state with
$\alpha=-\frac{2}{\delta}(1+3\pi(x_{10}-x_0));
\beta=-\frac{2}{\delta}(1+3\pi(x_{20}-x_0))$. The solution of
equation (8) after one period $T$ of the
oscillations in the linearized Poincar$\acute{e}$ map is given by
\begin{equation}
        Y(T) = Y(0)\exp(DG(V_{0})T)
\end{equation}
where $DG(V_{0})$ represents the time-independent
stability matrix of a periodic orbit connecting arbitrary
infinitesimal variations in the initial conditions $Y_{0}=Y(0)$ with
corresponding  change $Y(T)$ after one period $T$. The real parts of the
roots of the characteristic equation $det(DG(V_{0})-I\lambda)=0$, given by
\begin{equation}
        \lambda^4+A_3\lambda^3+A_2\lambda^2+A_1\lambda+A_0=0
\end{equation}
determines the stability of the periodic motion, with $A_3=2b+c$,
$A_2=b(b+c)-(\alpha+\beta)$, $A_1=-b(\beta+\alpha)+\alpha c$ and
$A_0=\alpha\beta$. Here $\lambda=(\lambda_j)$ represent the
eigenvalues of $DG(V_0)$. If one assumes $x_0 \approx x_{10} \approx x_{20}$, as in our numerical
experiment, then $\alpha=\beta=-\frac{2}{\delta}$. In general, complex
eigenvalues occur in complex conjugate pairs. Let us consider
$\alpha_k$, $\beta_k$ as the real and the imaginary parts of
$\lambda_k$, respectively ($\lambda_k=\alpha_k+i\beta_k$). If $\lambda_{k}$ is
real, the eigenvalues are simply the rate of contraction
($\alpha_k<0$) or expansion ($\alpha_k>0$) near the steady
state.  If $\lambda_k$ is complex, its real part $\alpha_k$ gives the rate of
contraction ($\alpha_k<0$) or expansion ($\alpha_k>0$) of the
spiral while its imaginary part $\beta_k$ contributes for the
frequency rotation. The eigenvalues of the linearized Poincar$\acute{e}$ map may thus be written as :
\begin{equation}
\sigma_k=\exp(\alpha_k T) \left(\cos(\beta_k T)+i\sin(\beta_k T) \right)
\end{equation}
It turns out from equation (12) that if $\alpha_k<0$ for all $\lambda_{k}$,
then all sufficiently small perturbations tend torward zero as $t\to\infty$ and
the steady state (node ({\it n}), saddle-node ({\it sn}), spiral ({\it
  sp})) is stable. If $\alpha_k>0$ for all $\lambda_k$, then any
small perturbation grows with time and the steady state ({\it n}, {\it sn}, {\it sp}) is unstable. In addition if there exist $m$ and $l$ such that
$\alpha_m<0$ and $\alpha_l>0$, the equilibrium state is unstable and is called a saddle. Having recourse to the above
analysis it follows that saddle-node, {\it sn} ($\lambda_k=+1$),
period-doubling, {\it pd} ($\lambda_k=-1$), Hopf ($\beta_k\neq 0$,
with $\alpha_k < 0$) and
symmetry-breaking ({\it sb}) bifurcations are expected to occur in the
coupled ratchets. Similar bifurcational scenario have been found in
periodically driven coupled single \cite{kozlowski95} and double
\cite{kenfack03} well Duffing oscillators.

\begin{figure}[h]
\vskip 0.25cm
\includegraphics[width=8.5cm,height=6cm]{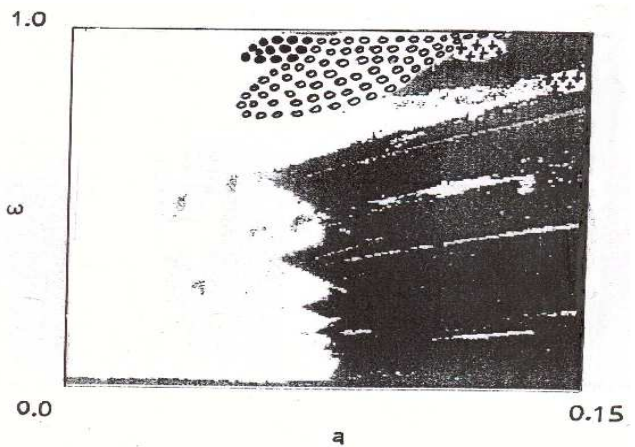}
\caption{\label{figure3} Two parameter phase diagram in the ($a
  -\omega$) plane. Regions of period-2 ($\circ$), period-3 ($\bullet$),
   period-4 ($+$), chaos (black) and non-attraction (white) can be identified.}
\end{figure}

\section{Bifurcation Diagrams} 
In order to investigate the dependence of the system on a single control parameter (in this case, the amplitude $a$ of the
        driving function), several bifurcation diagrams have been computed, some of which have been chosen to illustrate the
        general structure of the system. Each bifurcation diagram
        shows the projection of the attractors in the Poincar$\acute{e}$
        section onto the $x_2$ or the $v_2$ coordinates versus the
        control parameter.  We employ this technique in our numerical exploration using the period $T=2\pi/\omega$ of oscillation
        as the stroboscopic time; with the standard 4th order Runge
        Kutta algorithm.  

In what follows, we consider the bifurcational precedencies of
formation of the stable synchronized chaos in the
synchronization regime i.e in the neighbourhood of $0.89\le c \le
1.05$. We begin with a general overview of the behaviour of the two
coupled systems by displaying regions of existence of different
periodic and chaotic orbits in a two parameter phase diagram which is
plotted in ($a - \omega$) plane at fixed values of $b = 0.1$ and
$c=0.5$ (fig.~\ref{figure3}).  We obtained the parameter phase diagram
using the software dynamics~\cite{nusse98}. Different regions of  periodic
orbits of period-2 ($\circ$), period-3 ($\bullet$), period-4 ($+$) as well as
chaotic orbits (black region) are clearly visible. Besides the
periodic and the chaotic orbits, there also exist non-attracting
points (white region) for which the orbits diverge.

\begin{figure}[h]
\vskip 0.25cm
\includegraphics[width=8cm,height=12cm,angle=0]{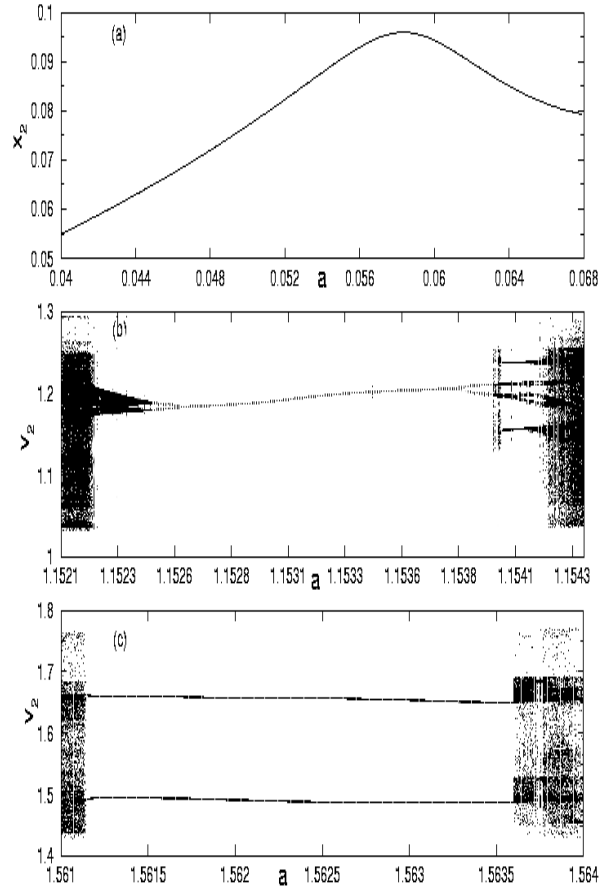}
\caption{\label{figure4} Bifurcation diagrams for the driving
  frequency $\omega = 0.3$, showing (a) a resonance near $a = 0.0559$,
  (b) a period-1 and (c) a large windows of period-2, both sandwiched in the
  chaotic regions.}
\end{figure}

\begin{figure}[h]
\vskip 0.25cm
\includegraphics[width=8cm,height=12cm,angle=0]{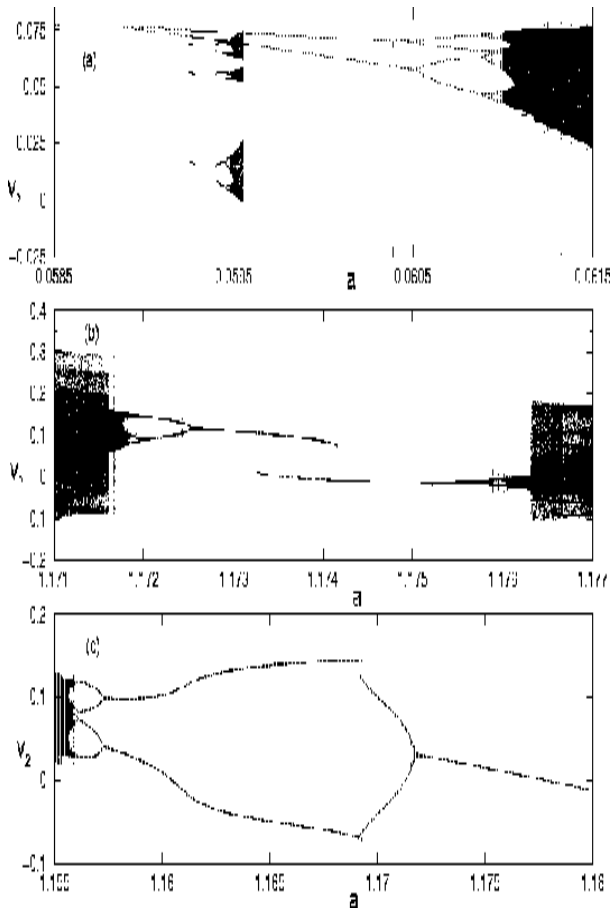}
\caption{\label{figure5} Bifurcation diagrams for the driving frequencies
  (a) $\omega = 0.4$, (b) $\omega = 0.5$ and (c) $\omega = 0.75$
  showing clearly the occurence of {\it sn}, {\it sb}, reversed {\it
  pd} and {\it pd} route to chaos.}
\end{figure}
In fig.~\ref{figure4} we show bifurcation diagrams for a comparatively small
driving frequency of $\omega=0.3$.  Different bifurcation sequences
occur as the driving amplitude $a$ is varied.  For instance in
fig.~\ref{figure4}(a), there is a resonance near $a=0.0559$ while a
period-1 attractor emerges in fig.~\ref{figure4}(b) from chaos,
through a reversed period-doubling ({\it pd}) and subsequently
follows the familiar {\it pd} route to chaos. For much larger amplitude of
the driving force, a large period-2 is suddenly created at around
$a=1.5612$ from a chaotic region, (see fig.~\ref{figure4}(c)). This period-2 attractor is finally
annihilated in a crisis event around $a=1.5638$ leading to a sudden chaotic
state (transcient chaotic state). After this chaotic state we find bubbles of bifurcation
dominating periodic windows (further, see fig.~\ref{figure7}(a) as a
prototype). 
When the driving frequency $\omega$ is increased, saddle-node ({\it sn}), symmetry breaking ({\it sb}) bifurcations, quasiperiodicity as
  well as period-doubling ({\it pd}) cascade predominate. For
  instance, fig.~\ref{figure5}(a) shows the occurence of {\it sn} in
  the vicinity of $a=0.05925$ and a {\it pd} cascade, at around $a=0.06075$, leading to chaos for $\omega=0.4$. Moreover $sb$ and {\it sn} resulting from a
  reversed {\it pd} are clearly visible in fig.~\ref{figure5}, (b) for
  $\omega=0.5$ and (c) for $\omega=0.75$. Similar routes identified in
  this transport model have also been found in the coupled Duffing model~\cite{kozlowski95,kenfack03}.

To complete our bifurcation study, we proceed to investigate large 
values of the driving frequency ($\omega>\omega_c=0.8$) and different ranges
of $a$. Here we find that the majority of the earlier observed
bifurcation scenario are also repeated except that for large $a$,
typically greater than the critical value $a_c=1.17$, periodic attractors
dominate. Fingerprints of such behaviour can readily be observed in
fig.~\ref{figure5}(c) for $\omega=0.75$, where a period-1 attractor is created via a {\it sb} bifurcation. 

\section{Characterizing Chaos}
\begin{figure}[h]
\vskip 0.25cm
\includegraphics[width=7cm,height=5cm,angle=0]{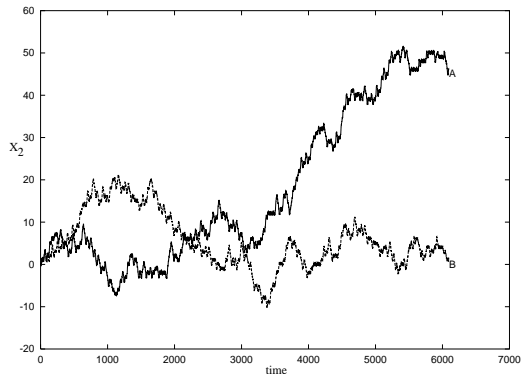}
\caption{\label{figure6} Typical chaotic trajectories of particle for two slightly different sets of initial conditions ($x_1=v_1=x_2=v_2=0.0$ for A and $x_1=v_1=v_2=0.0, x_2=0.01$ for B). This is a clear manifestation of SIC.}
\end{figure}

The best-known characteristic of a chaotic system is unpredictability,
        resulting from sensitivity to initial conditions (SIC). The slightest deviation
        from a given set of initial conditions yields a completely
        different output. In fig.~\ref{figure6}, two typical chaotic
        trajectories originated from two slightly different set of
        initial conditions ($x_2=0.0$ for A and $x_2=0.01$ for B) are
        displayed. The parameters have been selected from the chaotic
        region of the phase diagram of fig.~\ref{figure3} with the
        driving amplitude kept fixed $a=0.08092844$. Although both
        trajectories initially evolve together, after several time
        steps the state variables become essentially independent. This
        is a clear manifestation of the SIC, as witness of chaos.

Though SIC has come to be the hallmark of chaos, it is not unique to
        the chaotic regime. For example, a similar behaviour can be
        observed in the intermittency transition regime where the time series is nearly
        periodic. Thus, to better validate the results obtained above, it is
        necessary to investigate the dynamics of the ratchets in a
        more 
concrete space - the poincare cross-sections. Analysis of the dynamics
        in such space may help to eventually uncover the hidden
        structure in the behaviour of the coupled ratchets. In the poincare cross-section, all trajectories for
        a given set of control parameters converge to a single object (the attractor) in spite of SIC.
        The convergence to an attractor is guaranteed even if the control
        parameters are slightly varied.

\begin{figure}[h]
\vskip 0.25cm
\includegraphics[width=7cm, height=8cm]{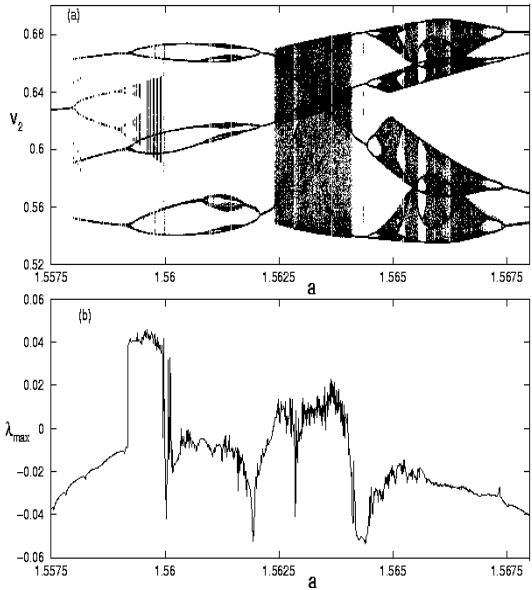}
\caption{\label{figure7} (a) Bifurcation diagram for $\omega=0.3$ and
  (b) the corresponding Lyapunov spectrum in the Poincare map.}
\end{figure}

        The set of Lyapunov exponents $\lambda_k$ provides an intuitively
        appealing and yet a very powerful measure of SIC and
        dissipation, both of which are required for a chaotic
        system. All $\lambda_k$ originates from linear stabilty
        analysis. In this approximation, all solutions are of the form
        $\exp(\lambda_kt)$, $k=1,2,...M$. The maximum Lyapunov
        exponent $\lambda_{max}$ is defined as:
\begin{equation}
 \lambda_{max}=\lim_{\tau\to+\infty}\frac{1}{\tau}\ln(||L(\tau)||),
\end{equation}
where $||L(\tau)||=(\delta x_1^2+\delta v_1^2+\delta x_2^2+\delta
        v_2^2)^{\frac{1}{2}}$ is obtained in the Poincar$\acute{e}$
        cross-section by solving numerically the variational equation (8) simultaneously with the systems (4) and (5).
        A positive Lyapunov exponent is a signature of chaos while zero and
        negative values of the exponent is an indication of a marginally stable
        or quasiperiodic orbit and periodic orbit, respectively. In
        fig.~\ref{figure7}(b), a spectrum of the $\lambda_{max}$,
        corresponding to the bifurcation diagram of
        fig.~\ref{figure7}(a) obtained for $\omega=0.3$, is displayed. Regions of chaotic and
        periodic solutions are well characterized. Various bifurcation
        scenario can be found including {\it sn}, {\it sb},
        quasiperiodic region and {\it pd} cascade leading to chaos. 
        Interestingly, bubbles of bifurcation beneath the chaotic
        region at $a=1.5630$, which by {\it eyes test} seems to be chaotic as
        well is not the
        case indeed. The Lyapunov spectrum shows that this regime is
        just a purely quasiperiodic one which yields precisely a
        period-4 attractor through a reversed {\it pd}.

\begin{figure}[h]
\vskip 0.25cm
\includegraphics[width=8cm,height=8cm,angle=0]{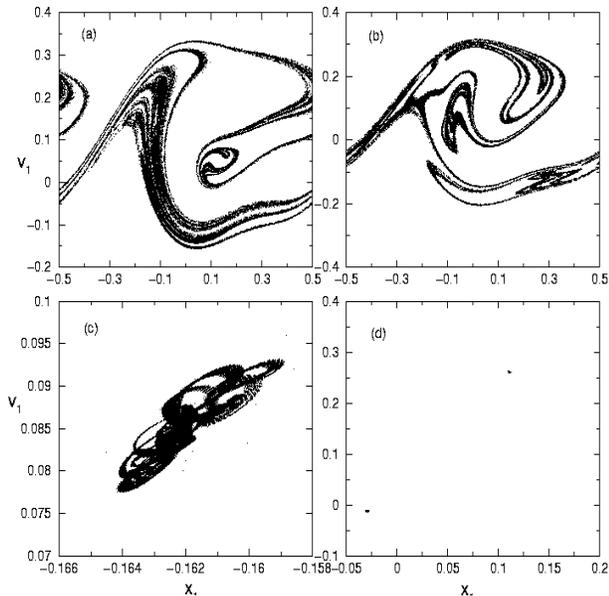}
\caption{\label{figure8} Typical attractors for parameters taken from fig.3, with a fixed value of the driving
  amplitude $a=0.08092844$. Shown are chaotic attractors (a) for
  $\omega=0.4$, (b) for $\omega =0.5$, (c) for $\omega=0.75$ and (d) a
  quasiperiodic attractor of period-2 for $\omega=0.8$.}
\end{figure}

Finally in fig.~\ref{figure8}, we visualise the attractors in the
Poincar$\acute{e}$ cross-sections for a fixed valued of the driving force
$a=0.080928844$ and for several values of the driving
frequency $\omega$ taken from fig.~\ref{figure3}. We find that as
$\omega$ is increased, the chaotic attractor shrinks (See
fig.~\ref{figure8}, (a) for $\omega=0.4$, (b) for
$\omega=0.5$ and (c) for $\omega=0.75$). As a
consequence, the attractor size gets enlarged as $\omega$
decreases. Essentially, this phenomenon has been attributed to the
collision of the attractor with a periodic orbit in the exterior of its basin and is thus referred to as boundary crises~\cite{ott02}. For larger $\omega$ values, quasiperiodic and periodic
        attractors were found to dominate. 
        As an example, fig.\ref{figure8}(d) shows a quasiperiodic
        attractor of period-2 for $\omega=\omega_c=0.8$.

\section{Conclusions}
In summary, we have investigated the dynamics of unidirectionally
        coupled deterministic ratchets and have shown varieties of
        bifurcation sequences including quasiperiodic route to
        chaos. Using the standard method of linear stability analysis,
        we have examined the stability of the steady state solution of
        the system leading to different types of bifurcations likely
        to occur in the neigbourhood of the synchronized region. For a
        given driving frequency, the bifurcations depend strongly on the
        values of the driving amplitude $a$ and are generally
        complicated. Besides quasiperiodicity, the familiar period-doubling and crisis route to chaos were also observed.
        In addition, the coupled ratchets exhibits symmetry-breaking
        bifurcations, resonance, saddle-node and bubbles of bifurcations. Using Lyapunov
        exponent spectrum as well as sensitivity to initial conditions, we
        characterized chaos in this system. A perusal of the Poincar$\acute{e}$ cross-sections revealed boundary crises in which
        the size of chaotic attractor is suddenly enlarged as the driving frequency
        is gradually decreased for a fixed driving amplitude. Although
        this system presents a very chaotic structure, it remains
        nevertheless ordered as the driving frequency
        becomes large, thereby providing critical parameters for a
        chaotic transport of particles. This makes the present study
        essentially interesting for technological applications.
\section{acknowledgements} 
UEV gratefully acknowledges Prof. Jose L. Meteos of the
Instituto de Fisica, Universidad Nacional Autonoma de Mexico, Mexico, for supplying relevant literatures and very useful discussions.
AK gratefully acknowledges the financial support of the Alexander von
Humboldt (AvH) Foundation/Bonn-Germany, under the grant of Research
fellowship no IV.4-KAM 1068533 STP. We are grateful to Dr. E. Arevalo
for a careful reading of the manuscript.


\begin{thebibliography}{99}
\bibitem{ruelle71}D.~Ruelle and F.~Takens, Commun.~Math.~Phys.~{\bf 20} (1971) 167
\bibitem{zhang02}Z.~Zhang, X.~Wag and M.~C.~Cross, Phys.~Rev.~E~{\bf 65} (2002) 056211
\bibitem{vincent04}U.~E.~Vincent, A.~N.~Njah, O.~Akinlade and A.~R.~T.~Solarin,
                Chaos~{\bf 14}, (2004) 1018.
\bibitem{vincent05}U.~E.~Vincent, A.~N.~Njah, O.~Akinlade and
  A.~R.~T.~Solarin, '{\it Phase synchronization in coupled chaotic ratchets} - Unpublished.
\bibitem{fujisaka83}H.~Fujisaka and T.~Yamada, Prog.~Theor.~Phys.~{\bf 69} (1983) 32
\bibitem{yamada83}T.~Yamada and H.~Fujisaka, Prog.~Theor.~Phys.~{\bf 70} (1983) 1240.
\bibitem{gauthier96}D.~J.~Gauthier and J.~C.~Bienfang, Phys.~Rev.~Lett.~{\bf 77} (1996) 1752.
\bibitem{ram00}R.J.~Ram, R.~Sporer, A.~R.~Blank and R.~A.~York, IEEE Trans.~Microwave Theor.~Techn.~{\bf 48} (2000) 1909.
\bibitem{zhang00}G.~H.~Y.~Zhang, H.~A.~Cerdeira, and
  S.~Chen, Phys.~Rev.~Lett.~{\bf 85} (2000) 3380.
\bibitem{ding96}M.~Ding and W.~Yang, Phys.~Rev.~E~{\bf 54} (1996) 2189.
\bibitem{kozlowski95}J.~Kozlowski, U.~Parlitz and W.~Lanterborn, Phys.~Rev.~E~{\bf 51} (1995) 1861.
\bibitem{kenfack03}A.~Kenfack, Chaos, Sol. and Fract.~{\bf 15} (2003) 205.
\bibitem{wirkus02}S.~Wirkus, R.~Rand, Nonl.~Dynamics~{\bf 30} (2002) 205.
\bibitem{woafo96}P.~Woafo, J.~C.~Chedjou and H.~B.~Fotsin, Phys.~Rev.~E~{\bf 54} (1996) 5929.
\bibitem{baker98}G.~L.~Baker, J.~A.~Blackburn, H.~J.~T.~Smith, Phys.~Rev.~Lett.~{\bf 81} (1998) 554.
\bibitem{chen03}Y.~Chen, G.~Rangarjan and M.~Ding, Phys.~Rev.E~{\bf 67} (2003) 026209.
\bibitem{hikihara01}T.~Hikihara, K.~Torii and Y.~Ueda, Phys.~Lett.~A.~{\bf 281} (2001) 155.
\bibitem{smith03}H. J.~T.~Smith, J.~A.~Blackburn and G.~L.~Baker,
  Int.~J.~Bifurc. Chaos~{\bf 13} (2003) 7.
\bibitem{kakis04}A.~F.~Va Kakis and R.~Rand, Int.~J.~of Nonlinear Mechanics~{\bf 39} (2004) 1079.
\bibitem{raj99}S. P.~Raj.~S.~Rajaskar and K.~Murali, Phys.~Lett.~A~{\bf 264}
                (1999) 283.
\bibitem{jung96}P.~Jung, J.~G.~Kissner, and P.~Hanggi,
  Phys.~Rev.~Lett.~{\bf 76} (1996) 3436.
\bibitem{mateos00}J.~L.~Meteos, Phys.~Rev.~Lett.~{\bf 84} (2000) 258.
\bibitem{mateos02}J.~L.~Meteos, Physica~D~{\bf 168} (2002) 205.
\bibitem{mateos03a}J.~L.~Meteos, Physica~A~{\bf 325} (2003) 92.
\bibitem{mateos03b}J.~L.~Meteos, Comm.~Nonl.~Sci.~Num.~Sim.~{\bf 8} (2003)
  253.
\bibitem{nusse98}H. E. Nusse and J.~A.~Yorke~,{\it Dynamics: Numerical Exploration}: Springer-Verlag, 1998.
\bibitem{ott02}E.~Ott,~{\it Chaos in Nonlinear Dynamical Systems}
  (Cambridge University Press, Cambridge, 2002).
\end{thebibliography}
\end{document}